\documentclass[12pt,preprint]{aastex}

\shorttitle{A 3D MHD Model of PNe}
\shortauthors{Tsui}
\begin{document}

\title{A Three-Dimensional Magnetohydrodynamic Model
\\ of Planetary Nebula Jets, Knots, and Filaments}
\author{K.H. Tsui}
\affil{Instituto de F\'{i}sica - Universidade Federal Fluminense
\\Campus da Praia Vermelha, Av. General Milton Tavares de Souza s/n
\\Gragoat\'{a}, 24.210-346, Niter\'{o}i, Rio de Janeiro, Brasil.}
\email{tsui$@$if.uff.br}
\date{}
\pagestyle{myheadings}
\baselineskip 18pt
	 
\begin{abstract}

The morphologies of planetary nebulae are believed to be
 self-organized configurations. These configurations are
 modeled by three-dimensional temporally self-similar
 magnetohydrodynamic solutions with radial flow, under
 the gravitational field of a central star of mass $M$.
 These solutions reproduce basic features, such as jets,
 point-symmetric knots, and filaments, through plasma
 pressure, mass density, and magnetic field lines.
 The time evolution function of the radial velocity
 starts as a slow wind and terminates as a fast wind.
   
\end{abstract}

\vspace{2.0cm}
\maketitle
\newpage
\section{Introduction}

The spherical observational features of planetary nebulae
 can be accounted for by the celebrated interacting wind
 model (Dyson and de Vries 1972, Kwok et al 1978),
 where the early slow wind of an intermediate mass star on the
 asymptotic giant branch phase is catched up by the following
 tenuous fast wind, generating the spherical features.
 The elliptic and bipolar features could also be reproduced
 by postulating the presence of a dense equatorial cloud
 (Kahn and West 1985, Mellema et al 1991),
 which has been imaged by high resolution instruments.
 Nevertheless, besides these spherical, elliptic, and bipolar
 features, images have also revealed point-symmetric knots
 (Miranda and Solf 1992, Lopez et al 1993, Balick et al 1993).
 These point-symmetric features plus the detection of magnetic
 fields in central stars (Jordan et al 2005) call for a
 magnetohydrodynamic (MHD) approach of planetary nebulae
 (Pascoli 1993, Chevalier and Luo 1994, Garcia-Segura 1997,
 Bogovalov and Tsinganos 1999, Matt et al 2000,
 Gardiner and Frank 2001).
 Recently, Tsui (2008) has regarded the morphologies of planetary
 nebulae as self-organized structures, described by temporally
 self-similar MHD solutions in spherical coordinates
 with radial velocity flow for explosions.
 The connection between the physical process of
 self-organization (Hasegawa 1985, Zhu et al 1995,
 Yoshida and Mahajan 2002, Kondoh et al 2004)
 and the mathematical analysis of self-similarity,
 using the Lagrangian radial label,
 has been discussed in detail.
 With axisymmetry, an equatorial plasma torus
 and a bipolar planetary nebula can be reproduced.
 In essence, current hydrodynamic and magnetofluid models
 attribute dynamic structures to the central star, or stars,
 as a cause to break the spherical isotropy,
 to account for the nebula morphologies, as a consequence.
 The self-similar approach relies on the global conservation
 properties of MHD plasma to reach self-organized configurations
 regardless of the initial conditions.
 The spherical isotropy of explosion is broken
 during the course of expansion because of the magnetic field,
 which is non-spherical by nature.
 
We follow this self-similar view that considers astrophysical
 phenomena with ejection origins as self-organized objects,
 whose configurations are solved by temporally self-similar
 MHD solutions
 (Low 1982a,b, Low 1984a,b, Osherovich et al 1993, 1995,
 Tsui and Tavares 2005, Tsui 2006, Tsui et al 2006).
 This includes active galactic nucleus (AGN) jets
 (Tsui and Serbeto 2007), classically treated as a
 steady state acretion-ejection MHD transport phenomenon
 from the acretion disk to the polar axis
 (Blandford and Payne 1982),
 and axisymmetric planetary nebulae.
 Here, we develope three-dimensional MHD solutions to describe
 additional planetary nebula features like jets (ansae), knots,
 and filaments. Unlike AGN jets that are extensive objects
 in galactic scales and are collimated by some physical mechanism,
 such as plasma pressure driven collimation (Tsui and Serbeto 2007),
 nebula jets are much smaller objects and are not collimated.
 We remark that three-dimensional self-similar MHD solutions
 are very rare. To our knowledge, there is only one published
 work which is devised specifically for interplanetary magnetic
 ropes (Gibson and Low 1998),
 which is not suitable for astrophysical phenomena.

\newpage
\section{Self-Similar Formulation}

The basic MHD equations in Eulerian fluid description are given by
\\
$${\partial\rho\over\partial t}+\nabla\cdot(\rho\vec v)\,
 =\,0\,\,\,,\eqno(1)$$

$$\rho\{{\partial\vec v\over\partial t}
 +(\vec v\cdot\nabla)\vec v\}\,
 =\,\vec J\times\vec B-\nabla p
 -\rho{GM\over r^{3}}\vec r\,\,\,,\eqno(2)$$

$${\partial\vec B\over\partial t}\,
 =\,-\nabla\times\vec E\,
 =\,\nabla\times(\vec v\times\vec B)\,\,\,,\eqno(3)$$

$$\nabla\times\vec B\,=\,\mu\vec J\,\,\,,\eqno(4)$$

$$\nabla\cdot\vec B\,=\,0\,\,\,,\eqno(5)$$

$${\partial\over\partial t}({p\over\rho^{\gamma}})
 +(\vec v\cdot \nabla)({p\over\rho^{\gamma}})\,=\,0\,\,\,.\eqno(6)$$
\\
Here, $\rho$ is the mass density, $\vec v$ is the bulk velocity,
 $\vec J$ is the current density, $\vec B$ is the magnetic field,
 $p$ is the plasma pressure, $\mu$ is the free space permeability,
 $\gamma$ is the polytropic index, and $M$ is the central mass
 which provides the gravitational field.

We consider a radially expanding plasma and seek self-similar
 solutions in time where the time evolution is described by the
 dimensionless evolution function $y(t)$.
 For this purpose, it is most convenient to think of Lagrangian
 fluid description, and consider the position vector of a given
 laminar flow fluid element $\vec r(t)$.
 Under self-similarity, the radial profile is time invariant
 in terms of the radial label $\eta=r(t)/y(t)$,
 which has the dimension of $r$.
 Physically, $\eta$ is the Lagrangian radial position
 of a fixed fluid element.
 With a finite plasma, the domain of $\eta$ is bounded by mass
 conservation
\\
$$0\,<\,\eta_{int}\,<\,\eta\,<\,\eta_{ext}\,\,\,.\eqno(7)$$
\\
As for the plasma velocity, we consider self-similar structures
 deriving from a spherically symmetric radial velocity which
 can be written as
\\
$$\vec v\,=\,{d\vec r(t)\over dt}\,
 =\,\{\eta{dy\over dt}, 0, 0\}\,
 =\,\{v, 0, 0\}\,\,\,.\eqno(8)$$
\\
Our self-similar parameter $\eta$, defined through
 the Lagrangian fluid label, explicitly represents
 the fluid velocity by the time evolution function $y(t)$.
 This evolution function will be solved self-consistently
 with respect to the spatial structures of the plasma.
 We emphasize that self-similarity, as a method, can be
 applied in different ways other than the one we use here.
 For example, Lou and his collaborators have treated an
 aggregating fluid under its self-gravitational field with a
 similarity variable $x=r(t)/at$, where $a$ is the sound speed,
 for isothermal fluid, $\gamma=1$,
 (Lou and Shen 2004, Bian and Lou 2005)
 and for a polytropic gas, $\gamma>1$,
 (Lou and Wang 2006, Lou and Gao 2006)
 to study relevant astrophysical phenomena.
 Extensions to a magnetofluid have been considered by Yu and Lou
 (2005) and by Lou and Wang (2007).
 Because of the linear dependence on time, this similarity
 variable $x$ refers to a reference frame moving at speed $a$,
 which is different from the radial plasma flow velocity $v$.
 Furthermore, different from our similarity variable $\eta$,
 $x$ here is not the Lagrangian label of a given fluid element.
 For this reason, the convective derivative remains explicit
 in the $x$ representation.
 As a result, the similarity variable of Lou amounts to finding
 the plasma structures in an adequate moving frame in the Eulerian
 $x$ fluid description.
 This resembles the analytic technique of going to a moving frame
 to look for stationary profile solutions for nonlinear phenomena
 such as nonlinear Alfven waves, solitons, etc.
 Because of this fundamentally different definition of
 self-similarity, the nature of the phenomena intended to
 describe is different.
 Our Lagrangian label self-similarity parameter is aimed to find
 spatial plasma configurations, and to determine the radial plasma
 flow velocity, consistent to the spatial configurations,
 through the evolution function.
 Since we are considering an isotropic radial plasma flow,
 a natural solution would be a hydrodynamic one-dimensional
 expanding plasma, with radially dependent mass density $\rho$
 and plasma pressure $p$, and with $\vec B=0$ and $\vec J=0$.
 Nevertheless, this one-dimensional hydrodynamic solution
 is highly unlikely because magnetic field fluctuations
 can be generated from current density fluctuations,
 even in the absence of a pre-existing magnetic field.
 With the magnetic fields, which are basically a two- or
 three-dimensional structure, coupling to the plasma
 will generate likewise two- or three-dimensional $\rho$ and $p$.

The independent variables are now transformed from $(r,\theta,\phi,t)$
 to $(\eta,\theta,\phi,y)$. We now proceed to determine the explicit
 dependence of $y$ in each one of the physical variables with
 this radial velocity using functional analysis.
 First, making use of Eq.(8), Eq.(1) renders
\\
$${\partial\rho\over\partial t}
 +{1\over r^2}{\partial\over\partial r}(r^2v\rho)\,
 =\,({\partial\rho\over\partial t}
 +v{\partial\rho\over\partial r})
 +\rho({\partial v\over\partial r}+{2v\over r})\,$$
 
$$ =\,{\partial\rho\over\partial y}{dy\over dt}
 +{3\rho\over y}{dy\over dt}\,
 =\,({\partial\rho\over\partial y}+{3\rho\over y})
 {dy\over dt}\,=\,0\,\,\,.\eqno(9a)$$
\\
To reach the second equality, we note that the first bracket
 in the first equality corresponds to the total time derivative
 of an Eulerian fluid element which amounts to the time derivative
 of a Lagrangian fluid element. As for the second bracket, it can
 be reduced by using $v=dr/dt=\eta dy/dt$ and
 $\partial v/\partial r=(1/y)(dy/dt)$. Solving this equation for
 the $y$ scaling by separating the time part gives
\\
$$\rho(\vec r,t)\,
 =\,{1\over y^3}\bar\rho(\eta,\theta,\phi)\,\,\,.\eqno(9b)$$
\\
As for Eq.(6), with $\alpha_{0}F=(p/\rho^{\gamma})$ where $\alpha_{0}$
 is a constant that carries the physical dimension so that $F$ is
 a dimensionless function, it follows
\\
$${\partial F\over\partial t}+v{\partial F\over\partial r}\,
 =\,{\partial F\over\partial y}{dy\over dt}\,
 =\,0\,\,\,,\eqno(10a)$$

$$({p\over\rho^{\gamma}})\,=\,\alpha_{0}F(\vec r,t)\,
 =\,{1\over y^0}\alpha_{0}\bar F(\eta,\theta,\phi)\,\,\,.\eqno(10b)$$
\\
As for Eq.(3), with the aid of Eq.(5), the magnetic fields are
\\
$${\partial B_{r}\over\partial t}
 +v{1\over r^{2}}{\partial\over\partial r}(r^{2}B_{r})\,
 =\,({\partial B_{r}\over\partial t}
 +v{\partial B_{r}\over\partial r})+{2v\over r}B_{r}\,$$
 
$$ =\,{\partial B_{r}\over\partial y}{dy\over dt}
 +{2B_{r}\over y}{dy\over dt}\,
 =\,({\partial B_{r}\over\partial y}+{2B_{r}\over y})
 {dy\over dt}\,
 =\,0\,\,\,,\eqno(11a)$$

$$B_{r}(\vec r,t)\,
 =\,{1\over y^2}\bar B_{r}(\eta,\theta,\phi)\,\,\,,\eqno(11b)$$

$${\partial B_{\theta}\over\partial t}
 +{1\over r}{\partial\over\partial r}( rvB_{\theta})\,
 =\,({\partial B_{\theta}\over\partial t}
 +v{\partial B_{\theta}\over\partial r})
 +{1\over r}B_{\theta}{\partial\over\partial r}(rv)\,$$
 
$$ =\,{\partial B_{\theta}\over\partial y}{dy\over dt}
 +{2B_{\theta}\over y}{dy\over dt}\,
 =\,({\partial B_{\theta}\over\partial y}+{2B_{\theta}\over y})
 {dy\over dt}\,
 =\,0\,\,\,,\eqno(12a)$$

$$B_{\theta}(\vec r,t)\,
 =\,{1\over y^2}\bar B_{\theta}(\eta,\theta,\phi)\,\,\,,\eqno(12b)$$

$${\partial B_{\phi}\over\partial t}
 +{1\over r}{\partial\over\partial r}( rvB_{\phi})\,
 =\,0\,\,\,,\eqno(13a)$$

$$B_{\phi}(\vec r,t)\,
 =\,{1\over y^2}\bar B_{\phi}(\eta,\theta,\phi)\,\,\,.\eqno(13b)$$
\\
Making use of Eq.(9b), we derive the plasma pressure from Eq.(10b)
\\
$$p(\vec r,t)\,
 =\,{1\over y^{3\gamma}}\alpha_{0}\bar F(\eta,\theta,\phi)
 \bar\rho^{\gamma}(\eta,\theta,\phi)\,
 =\,{1\over y^{3\gamma}}\alpha_{0}\bar F\bar\rho^{\gamma}\,
 =\,{1\over y^{3\gamma}}\bar p(\eta,\theta,\phi)\,\,\,.\eqno(14)$$
\\
Since Eq.(12a) and Eq.(13a) are of the same form, we conclude that,
 under self-similarity, $\bar B_{\phi}$ is a linear functional of
 $\bar B_{\theta}$ with
\\
$$\bar B_{\phi}\,=\,k\bar B_{\theta}\,\,\,.\eqno(15)$$
\\
Making use of Eq.(4) to eliminate the current density in Eq.(2),
 we get the momentum equation which has three components.
 The $\phi$, $\theta$, and $r$ components are respectively
\\
$$\bar B_{\theta}
 [{\partial\over\partial\theta}(k\bar B_{\theta}\sin\theta)
 -{\partial\bar B_{\theta}\over\partial\phi}]
 -\bar B_{r}
 [{\partial\bar B_{r}\over\partial\phi}
 -\sin\theta{\partial\over\partial\eta}(\eta k\bar B_{\theta})]
 -y^{4-3\gamma}{\partial\over\partial\phi}(\mu\bar p)\,
 =\,0\,\,\,,\eqno(16)$$

$$k\bar B_{\theta}
 [{\partial\over\partial\theta}(k\bar B_{\theta}\sin\theta)
 -{\partial\bar B_{\theta}\over\partial\phi}]
 +\bar B_{r}
 [\sin\theta{\partial\bar B_{r}\over\partial\theta}
 -\sin\theta{\partial\over\partial\eta}(\eta\bar B_{\theta})]
 +y^{4-3\gamma}\sin\theta{\partial\over\partial\theta}(\mu\bar p)\,
 =\,0\,\,\,,\eqno(17)$$

$$k\bar B_{\theta}
 [{1\over\eta\sin\theta}{\partial\bar B_{r}\over\partial\phi}
 -{1\over\eta}{\partial\over\partial\eta}(\eta k\bar B_{\theta})]
 -\bar B_{\theta}
 [{1\over\eta}{\partial\over\partial\eta}(\eta\bar B_{\theta})
 -{1\over\eta}{\partial\bar B_{r}\over\partial\theta}]
 -y^{4-3\gamma}{\partial\over\partial\eta}(\mu\bar p)\,$$
 
$$ =\,\mu\bar\rho y^2{d^2y\over dt^2}
 +\mu\bar\rho{GM\over\eta^2}
 \,\,\,.\eqno(18)$$
\\
We have reduced the general set of time-dependent ideal MHD
 equations, Eqs.(1-6), to a set of self-similar equations with
 appropriate time scalings, Eqs.(7-13).
 The general ideal MHD set has nonlinear terms of convective type
 $(\vec v\cdot\nabla)$.
 By using the fluid label description, the $(\vec v\cdot\nabla)$
 convective terms are absorbed in the Lagrangian time derivative
 representation.
 The structure of the nonlinear terms,
 absorbed in the Lagrangian fluid label formulation,
 will appear in the $\eta$ profile of the system.

\newpage
\section{Jet Structures}

After this self-similar formulation, we have to solve Eqs.(16-18)
 for the self-similar configurations.
 To solve these equations, we first separate the radial variable
 from the other two variables by writing
\\
$$\bar B_{r}(\eta,\theta,\phi)\,
 =\,A_{0}R(\eta)\tilde B_{r}(\theta,\phi)\,\,\,,\eqno(19)$$
\\
and likewise for $\bar B_{\theta}$,
 where $\bar B_{\phi}=k\bar B_{\theta}$.
 Furthermore, we take
\\
$$\bar p(\eta,\theta,\phi)\,
 =\,p_{0}R^2(\eta)\tilde p(\theta,\phi)\,\,\,.\eqno(20)$$
\\
Here, $R(\eta)$, $\tilde B_{r}(\theta,\phi)$, and
 $\tilde p(\theta,\phi)$ are dimensionless functions,
 and $A_{0}$ and $p_{0}$ carry the dimensions of magnetic
 field and pressure respectively.
 We have taken an $R^2(\eta)$ dependence because
 plasma pressure is a quadratic positive quantity.
 We take $A_{0}=1$ for unit amplitude magnetic fields,
 such that $p_{0}$ is relative to this amplitude.
 Specifically, we take
\\
$$R(\eta)\,=\,(a\eta)^{-n}\,\,\,,\eqno(21)$$
\\
to represent a power law decaying field with distance,
 where $a$ is a normalizing parameterof $\eta$.
 Considering $(4-3\gamma)=0$, Eqs.(16,17) are respectively
\\
$$\tilde B_{\theta}
 [{\partial\over\partial\theta}(k\tilde B_{\theta}\sin\theta)
 -{\partial\tilde B_{\theta}\over\partial\phi}]
 -\tilde B_{r}(n-1)\sin\theta k\tilde B_{\theta}
 -{\partial\over\partial\phi}
 ({1\over 2}\tilde B^2_{r}+\mu p_{0}\tilde p)\,
 =\,0\,\,\,,\eqno(22)$$

$$k\tilde B_{\theta}
 [{\partial\over\partial\theta}(k\tilde B_{\theta}\sin\theta)
 -{\partial\tilde B_{\theta}\over\partial\phi}]
 +\tilde B_{r}(n-1)\sin\theta \tilde B_{\theta}
 +\sin\theta{\partial\over\partial\theta}
 ({1\over 2}\tilde B^2_{r}+\mu p_{0}\tilde p)\,
 =\,0\,\,\,,\eqno(23)$$
\\
We now separate the azimuthal dependence by writing
\\
$$\tilde B_{r}(\theta,\phi)\,
 =\,\Theta_{r}(\theta)\Phi(\phi)\,\,\,,\eqno(24a)$$
 
$$\tilde B_{\theta}(\theta,\phi)\,
 =\,\Theta(\theta)\Phi(\phi)\,\,\,,\eqno(24b)$$

$$\tilde p(\theta,\phi)\,
 =\,\Theta^2_{p}(\theta)\Phi^2(\phi)\,\,\,,\eqno(24c)$$
\\
where $\tilde B_{\phi}=k\tilde B_{\theta}$. These functional 
 dependences give Eq.(22) as
\\
$$\Theta{\partial\over\partial\theta}(k\Theta\sin\theta)
 -\Theta_{r}(n-1)\sin\theta k\Theta\,
 =\,[\Theta^2-(\Theta^2_{r}+2\mu p_{0}\Theta^2_{p})]
 {1\over\Phi}{\partial\Phi\over\partial\phi}
 \,\,\,.$$
\\
Considering the separation constant $im$, such that
\\
$${\partial\Phi\over\partial\phi}\,=\,im\Phi\,\,\,,\eqno(25a)$$

$$\Phi(\phi)\,=\,e^{+im\phi}\,
 =\,\cos m\phi + i\sin m\phi\,\,\,,\eqno(25b)$$
\\
we then have
\\
$$\Theta{\partial\over\partial\theta}(\Theta\sin\theta)
 -\Theta_{r}(n-1)\sin\theta\Theta\,
 =\,+{i\over k}m[\Theta^2-(\Theta^2_{r}+2\mu p_{0}\Theta^2_{p})]
 \,\,\,,\eqno(26)$$
\\
which identifies $k=+i$. Following the same procedures,
 Eq.(23) reads
\\
$$\Theta{\partial\over\partial\theta}(\Theta\sin\theta)
 -\Theta_{r}(n-1)\sin\theta\Theta\,
 =\,+m\Theta^2
 +{1\over 2}\sin\theta{\partial\over\partial\theta}
 (\Theta^2_{r}+2\mu p_{0}\Theta^2_{p})
 \,\,\,.\eqno(27)$$
\\
The left sides of these two equations are the same,
 which allows the right sides be equated to give
\\
$$\sin\theta{\partial\Theta^2_{*}\over\partial\theta}\,
 =\,-2m\Theta^2_{*}\,\,\,,\eqno(28a)$$
 
$$\Theta^2_{*}\,=\,(\Theta^2_{r}+2\mu p_{0}\Theta^2_{p})
 \,\,\,.\eqno(28b)$$
\\
To solve for $\Theta^2_{*}$, we integrate Eq.(28a) to get
\\
$$\ln \Theta^2_{*}\,
 =\,-2m\int {d\theta\over\sin\theta}\,
 =\,-2m\int {\sin\theta d\theta\over\sin^2\theta}\,
 =\,+\ln ({1+x\over 1-x})^m\,\,\,,\eqno(29)$$
\\
where we have multiplied and divided the right side by
 $\sin\theta$ to implement the integration, and $x=\cos\theta$.
 This solution of $\Theta^2_{*}$ is singular at $x=+1$
 for positive $m$, and $x=-1$ for negative $m$.
 Such solution gives jet features on magnetic field lines
 and plasma density.

\newpage
\section{Special n=2 Case}

To get $\Theta$, instead of solving either Eq.(26) or Eq.(27),
 we make use of Eq.(5) to get
\\
$$\nabla\cdot\vec B\,
 =\,{1\over y^3}({1\over\eta}R(\eta))
 \{-(n-2)\tilde B_{r}+{1\over\sin\theta}
 [{\partial\over\partial\theta}(\sin\theta\tilde B_{\theta})
 +{\partial\over\partial\phi}(k\tilde B_{\theta})]\}\,
 =\,0\,\,\,,\eqno(30a)$$

$${\partial\over\partial\theta}(\sin\theta\Theta)\,
 =\,m\Theta+(n-2)\sin\theta\Theta_{r}\,\,\,.\eqno(30b)$$
\\
Subsituting Eq.(30b) into Eq.(26) and Eq.(27) respectively gives
\\
$$\sin\theta\Theta\Theta_{r}\,=\,m\Theta^2_{*}\,\,\,,\eqno(31a)$$

$$2\Theta\Theta_{r}\,
 =\,-{\partial\over\partial\theta}\Theta^2_{*}\,\,\,.\eqno(31b)$$
\\
It can be shown readily that these two equations are consistent
 to Eq.(28a). We note that Eq.(30b) couples $\Theta$ with
 $\Theta_{r}$. To decouple these two functions, we consider
 the special, but probably practical, case of
\\
$$n\,=\,2\,\,\,.\eqno(32)$$
\\
Multiplying over by $\sin\theta$ and defining
 $P(\theta)=\sin\theta\Theta$, Eq.(30b) can be integrated to give
\\
$$P(x)\,=\,({1-x\over 1+x})^{m/2}\,\,\,,\eqno(33a)$$

$$\Theta(x)\,=\,{(1-x)^{(m-1)/2}\over (1+x)^{(m+1)/2}}
 \,\,\,,\eqno(33b)$$

$$\Theta_{r}(x)\,=\,m({1+x\over 1-x})^{3m/2}
 \,\,\,,\eqno(33c)$$

$$2\mu p_{0}\Theta^2_{p}(x)\,
 =\,[({1+x\over 1-x})^{m}-m^2({1+x\over 1-x})^{3m}]\,
 >\,0\,\,\,,\eqno(33d)$$
\\
where Eq.(33c) is obtained from Eq.(31a), and $\Theta^2_{p}(x)$
 can be recovered from $\Theta^2_{*}(x)$ of Eq.(28b).
 From these solutions, we see that $B_{r}$ and $p$ are singular
 at $x=+1$ through $\Theta_{r}(x)$ and $\Theta^2_{p}(x)$,
 whereas $B_{\theta}$ and $B_{\phi}$ are singular
 at $x=-1$ through $\Theta(x)$. 
 We require the singularities be integrable in $x$,
 which demands the power of the singularities be less than unity.
 By inspection of the terms, we conclude that $m<1/3$.
 Let us take
\\
$$m\,=\,{1\over 4}\,\,\,,\eqno(34)$$
\\
to show $\Theta^2_{*}(x)$ in Fig.1 in a polar plot,
 $\Theta_{r}(x)$ and $\Theta(x)$ in Fig.2 and Fig.3
 respectively.
 Figures 1 and 2 show an integrable singularity at $x=+1$,
 or $\theta=0$.
 Furthermore, Fig.3 shows a weaker singularity at $x=+1$
 than the one of Fig.2.
 However, Fig.3 shows another singularity at $x=-1$,
 while Fig.2 is regular at that location.
 As for $\Theta^2_{p}(x)$, the second term of Eq.(33d)
 exceeds the first term when
\\
$$m({1+x\over 1-x})^m\,>\,1\,\,\,,$$
\\
or as $x$ gets very close to unity.
 Writing $x=(1-\delta)$, we get $(\delta/2)^m<m$,
 or $(\delta/2)<m^{1/m}$.
 With Eq.(34), we have $\delta=(1/2)^7$.
 Consequently, $\Theta^2_{p}(x)$ gets smaller as $x$
 approaches unity, and it vanishes at $x=(1-\delta)$,
 beyond this point, it gets negative.
 To interpret this negative plasma pressure amplitude,
 $\Theta^2_{p}(x)$ now points backwards to the $x=-1$
 direction in a polar plot,
 giving a jet structure as in Fig.4.
 This jet structure is along the radial magnetic field
 given by $\Theta_{r}(x)$.
 
As for the radial component, Eq.(18), with $(4-3\gamma)=0$,
 it reads
\\
$$\mu\bar\rho\eta y^2{d^2y\over dt^2}
 +\mu\bar\rho{GM\over\eta^2}\,
 =\,({1\over\eta}R^2)\tilde B_{\theta}
  [k{1\over\sin\theta}{\partial\tilde B_{r}\over\partial\phi}
 +{\partial B_{r}\over\partial\theta}]
  +2n({1\over\eta}R^2)(\mu p_{0}\tilde p)\,$$

$$=\,({1\over\eta}R^2)\tilde B_{\theta}
  [-m\Theta_{r}+\sin\theta{\partial\Theta_{r}\over\partial\theta}]
  \Phi
  +2n({1\over\eta}R^2)\mu p_{0}(\Theta_{p}\Phi)^2\,\,\,.\eqno(35)$$
\\
With $\alpha$ as the separation constant, we have
\\
$$({1\over\eta}R^2)
 \{\tilde B_{\theta}
  [-m\Theta_{r}+\sin\theta{\partial\Theta_{r}\over\partial\theta}]
  \Phi
  +2n\mu p_{0}(\Theta_{p}\Phi)^2\}\,
 =\,\mu\bar\rho({GM\over\eta^2}+\alpha\eta)\,\,\,,\eqno(36a)$$

$${d^2y\over dt^2}\,=\,{\alpha\over y^2}\,\,\,.\eqno(36b)$$
\\
We write the mass density as
\\
$$\bar\rho(\eta,\theta,\phi)\,
 =\,\rho_{0}\textbf{R}(\eta)\tilde\rho(\theta,\phi)
 \,\,\,,\eqno(37)$$
\\
where $\rho_{0}$ carries the dimension of mass density,
 and $\textbf{R}(\eta)$ and $\tilde\rho(\theta,\phi)$
 are dimensionless functions.
 We can identify immediately from Eq.(35a) that
\\
$$\textbf{R}(\eta)\,
 =\,{R^2\over (GM/\eta+\alpha\eta^2)}\,\,\,,\eqno(38a)$$
 
$$\mu\rho_{0}\tilde\rho\,
 =\,\{\tilde B_{\theta}
  [-m\Theta_{r}+\sin\theta{\partial\Theta_{r}\over\partial\theta}]
  \Phi
  +2n\mu p_{0}(\Theta_{p}\Phi)^2\}\,
 =\,[2n\mu p_{0}\Theta^2_{p}-2m\Theta\Theta_{r}]\Phi^2
 \,\,\,.\eqno(38b)$$
\\
From the first of these two equations, we deduce that $\alpha$
 is positive, such that $\textbf{R}(\eta)$ is analytic.
 As for the second equation, it reads
$$\mu\rho_{0}\tilde\rho\,
 =\,\{n[({1+x\over 1-x})^{m}-m^2({1+x\over 1-x})^{3m}]
 -2m^2{(1+x)^{(2m-1)/2}\over (1-x)^{(2m+1)/2}}\}\Phi^2\,
 >\,0\,\,\,.\eqno(39)$$
\\
This mass density has a singularity at $x=+1$.
With $m=1/4$, we note that the numerator of the last term
 in Eq.(38), $(1+x)^{(2m-1)/2}$, has a negative power.
 This gives a singularity at $x=-1$. Since the power of
 this singularity is $-1/4$, it is also integrable as well.
 The mass density distribution is shown in Fig.5
 with also a jet structure in the $x=-1$ direction,
 because of the radial magnetic field.
 If we consider $m<0$ negative,
 we should point out that Figs.(1-5) would turn upside down.
 Consequently, the jet structures would be on both sides
 of the polar axis.
 Some examples of these jets appear in M2-9 Twin Jet Nebula,
 CRL 2688 Egg Nebula, NGC 3242, NGC 6826, NGC 7009.
 As for the time evolution function of Eq.(35b), we multiply
 over by $dy/dt$ to get the first integral as
\\
$$({dy\over dt})^2\,
 =\,2(H-{\alpha\over y})\,>\,0\,\,\,,\eqno(40a)$$
\\
where $H$ is an integration constant. Knowing that $\alpha$
 is positive, and $y(0)=1$ by definition of Lagrangian
 fluid label, and the right side of this equation
 has to be positive, we conclude that $H>\alpha>0$.
 This gives the plasma wind velocity a slow start initially
 with $(dy/dt)^2=2(H-\alpha)$ which evolves into a fast
 terminal wind of
\\
$$({dy\over dt})^2\,=\,2H\,\,\,.\eqno(40b)$$
\\

\newpage
\section{Knot and Filament Structures}

With the momentum equation solved, the magnetic fields
 are given by
\\
$$\bar B_{r}\,=\,+R(\eta)\Theta_{r}\cos m\phi\,\,\,,\eqno(41a)$$ 

$$\bar B_{\theta}\,=\,+R(\eta)\Theta\cos m\phi\,\,\,,\eqno(41b)$$ 

$$\bar B_{\phi}\,=\,-R(\eta)\Theta\sin m\phi\,\,\,.\eqno(41c)$$ 
\\
The magnetic field lines are given by
\\
$${\Theta_{r}\cos m\phi\over d\eta}\,
 =\,{\Theta\cos m\phi\over\eta d\theta}\,
 =\,-{\Theta\sin m\phi\over\eta\sin\theta d\phi}
 \,\,\,,\eqno(42a)$$
\\
which can be written as
\\
$${d\eta\over\eta}\,
 =\,-m{(1+x)^{2m}\over (1-x)^{2m}}dx\,
 =\,-m{(1+x)^{2m+1}\over (1-x)^{2m-1}}{d\phi\over\tan m\phi}
 \,\,\,.\eqno(42b)$$
\\
With $\eta_{0}$ as an integration constant, the first equality gives
\\
$$\ln{\eta\over\eta_{0}}\,
 =\,-m\int {(1+x)^{2m}\over (1-x)^{2m}} dx
 \,\,\,,\eqno(43a)$$
\\
which can be integrated numerically, as is shown in Fig.6.
 Since the singularity is integrable, $\eta$ is finite at $x=+1$.
 The second equality corresponds to
\\
$${d\theta\over\sin\theta}\,
 =\,-{ d\phi\over\tan m\phi}\,\,\,,$$
\\
and can be integrated to give
\\
$${(1+x)^{m}\over (1-x)^{m}}\,
 =\,K(\sin m\phi)^2\,\,\,,\eqno(43b)$$
\\
where $K$ is an integration constant.
 With the interval of $x$ between $(-1,+1)$, or $\theta$
 between $(\pi,0)$, the left side of Eq.(43b),
 ${(1+x)^{m}/(1-x)^{m}}$ labelled on the left axis,
 covers an interval $(0,\infty)$ with positive $m$,
 and is plotted in Fig.7 against $x$ between $(-1,+1)$
 labelled on the bottom axis.
 The right side, $K(\sin m\phi)^2$ labelled on the right axis,
 is also plotted against $m\phi$ on the top axis,
 with a scale between $0$ and $\pi/2$, in the same figure.
 In order to view the mapping
 between the right side and the left side,
 we assign a large constant $K$.
 With $m=1/4$ and $K=3$, as $x$ departs from $-1$,
 or $\theta$ from $\pi$, $m\phi$ departs from $0$,
 and it maps a root of $\phi$.
 As $x$ approaches $+1$, or $\theta$ approaches $0$,
 $m\phi$ reaches $\pi/2$, which takes $\phi$ over $(0,2\pi)$.
 On the return path of the field lines,
 $x$ decreases from $+1$ back to $-1$,
 bringing $m\phi$ from $\pi/2$ to $\pi$
 along the descending branch of $K(\sin m\phi)^2$,
 which takes $\phi$ over $2\pi,4\pi$.
 This descending branch, which is the continuation of Fig.7,
 is not shown.
 To summarize, the field lines starting at $x=-1$ and $\eta=+1$
 go through the $x=(-1,+1,-1)$ cycle once,
 with $\eta=(+1,+2.3,+1)$,
 while completing the $m\phi=(0,\pi)$ cycle once,
 covering $\phi=(0,4\pi)$, before closing on themselves again.
 This generates helical field lines, as shown in Fig.8,
 on the surface of revolution of Fig.6.
 Since the singularities at $x=\pm 1$ are integrable,
 and also because of the circulating nature of
 the fields $\bar B_{\theta}$ and $\bar B_{\phi}$,
 the magnetic field lines converge to $\eta=1$ at $x=-1$
 axis and to $\eta=2.3$ at $x=+1$ axis, as shown in Fig.6.
 These locations correspond to point-symmetric magnetic
 knots, where the field strength is infinite.
 As for the filaments, they correspond to the helical
 magnetic field lines in space, as shown in Fig.8.
 These same field lines give a different shape
 when they are viewed at different orientations,
 such as in Fig.9 and Fig.10.
 Further field lines can be generated with, for example,
 $m=1/3.5=2/7=4/14$.
 In this case, the field lines starting at $x=-1$ and $\eta=+1$
 go through the $x=(-1,+1,-1)$ cycle once,
 with $\eta=(+1,+2.3,+1)$,
 while completing the $m\phi=(0,\pi/2,\pi)$ cycle once,
 covering $\phi=(0,7\pi/4,7\pi/2)$.
 Since the lowest $2\pi$ multiple of the $\phi$ cycle is four,
 the field lines have to complete four cycles of $x$ and
 of $m\phi$ to make $\phi$ covering $(0,7\pi,14\pi)$,
 such that the field lines can close on themselves again.
 With different values of $\eta_{0}$ of Eq.(43a),
 we can fill up the space with shells of field lines
 of some $m$ less than 1/3 up to $\eta_{ext}$ of Eq.(7),
 and dot the polar axis with a line of magnetic knots.
 Despite the radial component at $x=+1$ giving the plasma
 jet structures, the field lines are closed at $\eta=2.3$
 because of the circulating nature of the meridian and
 azimuthal components.
 With $m<0$ negative, the mirror images of the knots and
 field lines can be superimposed on those with $m>0$ positive,
 generating lines of knots and concentric shells
 of magnetic field lines.
 Examples of knots can be found in M2-9 Twin Jet Nebula,
 NGC 5307, and filaments in MyCn 18 Hourglass Nebula,
 NGC 6543 Cat's Eye Nebula, NGC 2392 Eskimo Nebula,
 M2-9 Twin Jet Nebula, NGC 6543.

\newpage
\section{General n Case}

We now solve Eq.(30b) for an arbitrary $n$.
 Combining Eq.(30b) an Eq.(31a), we get
\\
$$(1-x^2){d\over dx}P^2(x)+2mP^2(x)\,
 =\,-2m(n-2)(1-x^2)\Theta^2_{*}(x)\,\,\,,\eqno(44a)$$
\\
where $P(x)=(1-x^2)^{1/2}\Theta(x)$ and $\Theta^2_{*}(x)$
 is given by Eq.(29). If the boundary condition is known,
 this equation can be integrated numerically to get
\\
$$P^2(x)\,
 =\,-2m\int^{x}_{-1}[{P^2(x)\over (1-x^2)}
 +(n-2)\Theta^2_{*}(x)]dx
 +P^2(-1)\,\,\,.\eqno(44b)$$
\\
To obtain the boundary condition at $x=-1$, we note that
 the right side of Eq.(44a) is equal to
 $-2m(n-2)(1+x)^{m+1}/(1-x)^{m-1}$.
 Since $m<1$ positive, this term vanishes at $x=-1$ and
 at $x=+1$ as well.
 As a result, in the neighborhood of $x=\pm 1$, $P^2(x)$
 is described by the homogeneous version of Eq.(44a),
 where the right side is null. The homogeneous solution
 can be solved readily as
\\
$$P^2(x)\,=\,({1-x\over 1+x})^{m}\,\,\,,\eqno(45)$$
\\
which provides the needed boundary condition.
 The self-similar functions are, therefore, given by
\\
$$\Theta(x)\,=\,{P(x)\over\sin\theta}
 \,\,\,,\eqno(46a)$$

$$\Theta_{r}(x)\,=\,{m\Theta^2_{*}(x)\over P(x)}
 \,\,\,,\eqno(46b)$$

$$2\mu p_{0}\Theta^2_{p}(x)\,
 =\,[\Theta^2_{*}(x)-\Theta^2_{r}(x)]\,
 >\,0\,\,\,,\eqno(46c)$$
\\
As for the magnetic field lines, they are described by
\\
$$\ln{\eta\over\eta_{0}}\,
 =\,-m\int {\Theta^2_{*}(x)\over P^2(x)} dx\,\,\,,\eqno(47a)$$
 
$${d\theta\over\sin\theta}\,
 =\,-{ d\phi\over\tan m\phi}\,\,\,.\eqno(47b)$$
\\
The second equation remains the same, while the first equation
 can be integrated numerically to get the field lines.
 With $n=3$, the corresponding self-similar functions
 of Figs.2-6 are evaluated anew, and are presented in Figs.11-15
 respectively. From Fig.12 and Fig.15, we can see that the
 source term in Eq.(44a) makes $\Theta(x)$ and the $(\eta-x)$
 mapping more symmetric. The corresponding magnetic field lines
 are also shown in Figs.16-18, with knots, $x=\pm 1$ and $\eta=+1$,
 at equal distance from the center.

To conclude, we have regarded the morphologies of planetary
 nebulae as self-organized configurations. These configurations
 are modeled by temporally self-similar MHD solutions.
 To complement an earlier publication (Tsui 2008) for axisymmetric
 features, we have presented a three-dimensional self-similar
 model with $\gamma=4/3$,
 which reproduces features like jets, point-symmetric knots,
 and filaments, through plasma pressure, mass density, and
 magnetic field lines.
 The time evolution function of the self-similar solutions
 starts the plasma expansions as a slow wind, and terminates
 as a fast wind.
 With this three-dimensional model, which completes the
 earlier two-dimensional axisymmetric model, we have covered
 most of the existing features in planetary nebulae with this
 self-similar approach for self-organized configurations.
 Considering that the extragalactic AGN polar jets
 could be accounted for on the same ejection basis
 (Tsui and Serbeto 2007),
 through a plasma pressure driven collimation process,
 which differs from the classical accretion-ejection spatially
 self-similar steady state MHD transport model
 (Blandford and Payne 1982),
 we believe temporally self-similar MHD configurations are
 universal manifestations of self-organized astrophysical
 ejection phenomena.

\centerline{\bf Acknowledgments}

The author is deeply grateful to Dr. B.C. Low for the inspiring
 thoughts and physical insights of self-similar solutions,
 and to Prof. Akira Hasegawa for the very essential concept
 of self-organization in fluids and plasmas.

\clearpage
\begin{figure}
\plotone{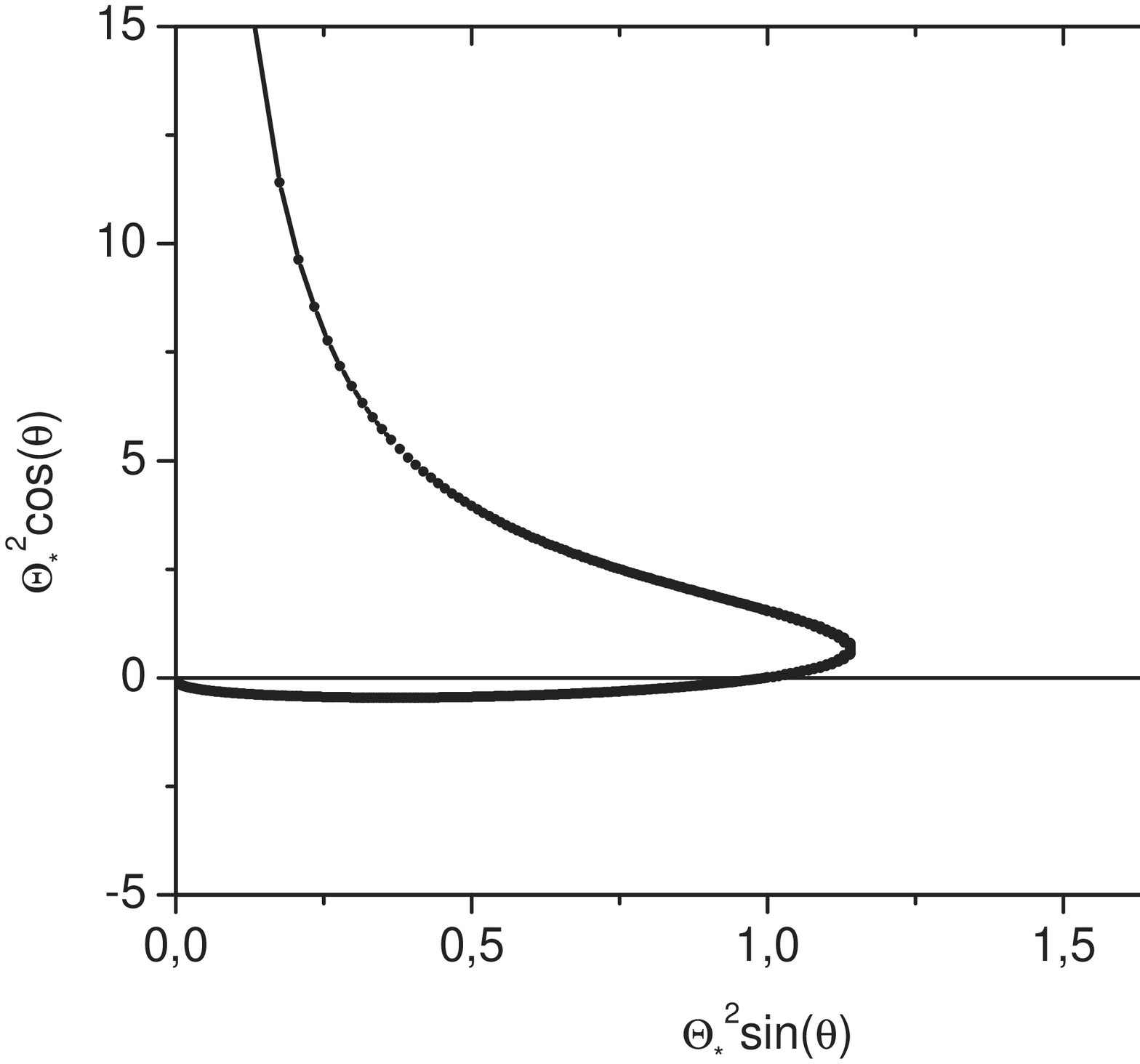}
\caption{The function $\Theta^{2}_{*}(x)$ with $x=\cos\theta$
 is shown in a polar plotted indicating an integrable singularity
 at $x=+1$ for a jet structure.}
\label{fig.1}
\end{figure}

\clearpage
\begin{figure}
\plotone{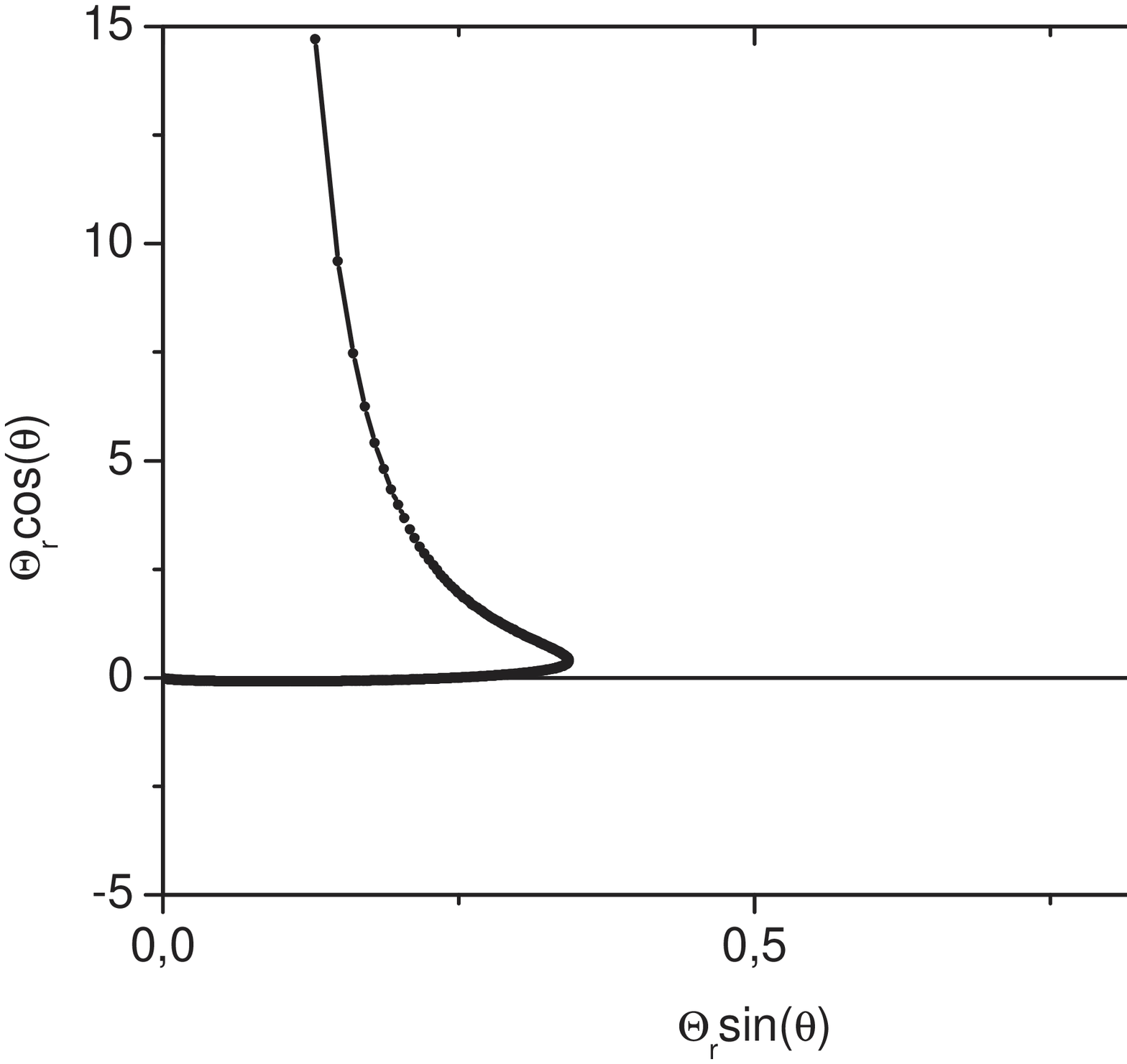}
\caption{The function $\Theta_{r}(x)$ with $x=\cos\theta$
 is shown in a polar plotted indicating an integrable singularity
 at $x=+1$ for a jet structure.}
\label{fig.2}
\end{figure}

\clearpage
\begin{figure}
\plotone{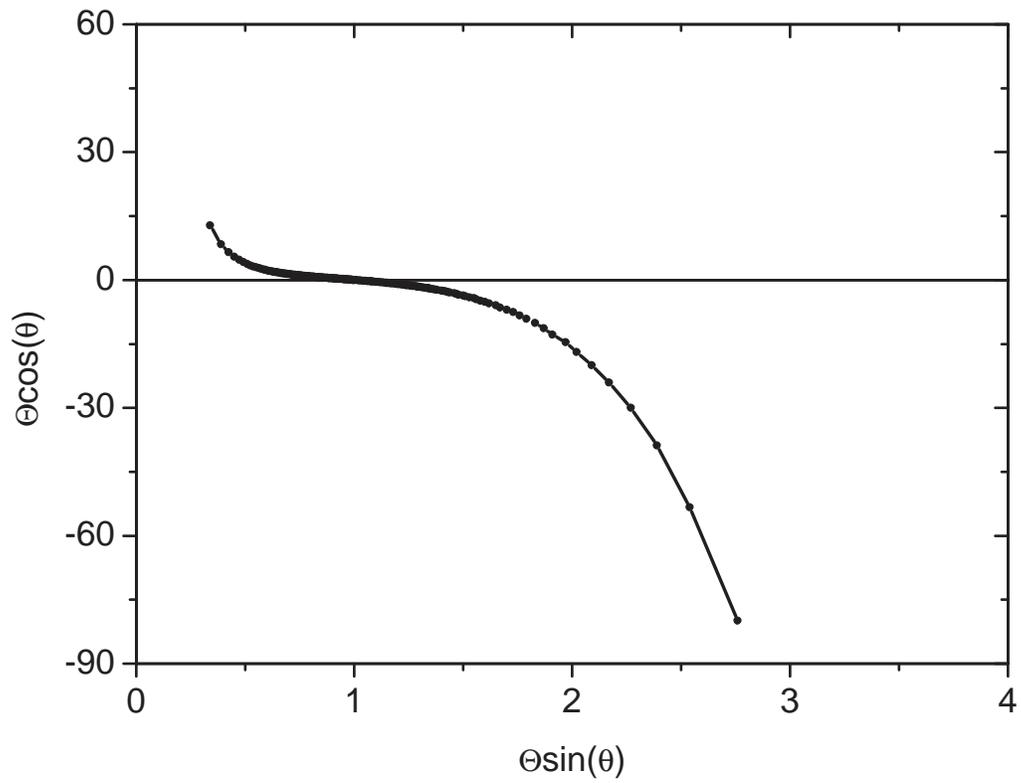}
\caption{The function $\Theta(x)$ with $x=\cos\theta$
 is shown in a polar plotted indicating integrable singularities
 at $x=+1$ and at $x=-1$ for knot structures.}
\label{fig.3}
\end{figure}

\clearpage
\begin{figure}
\plotone{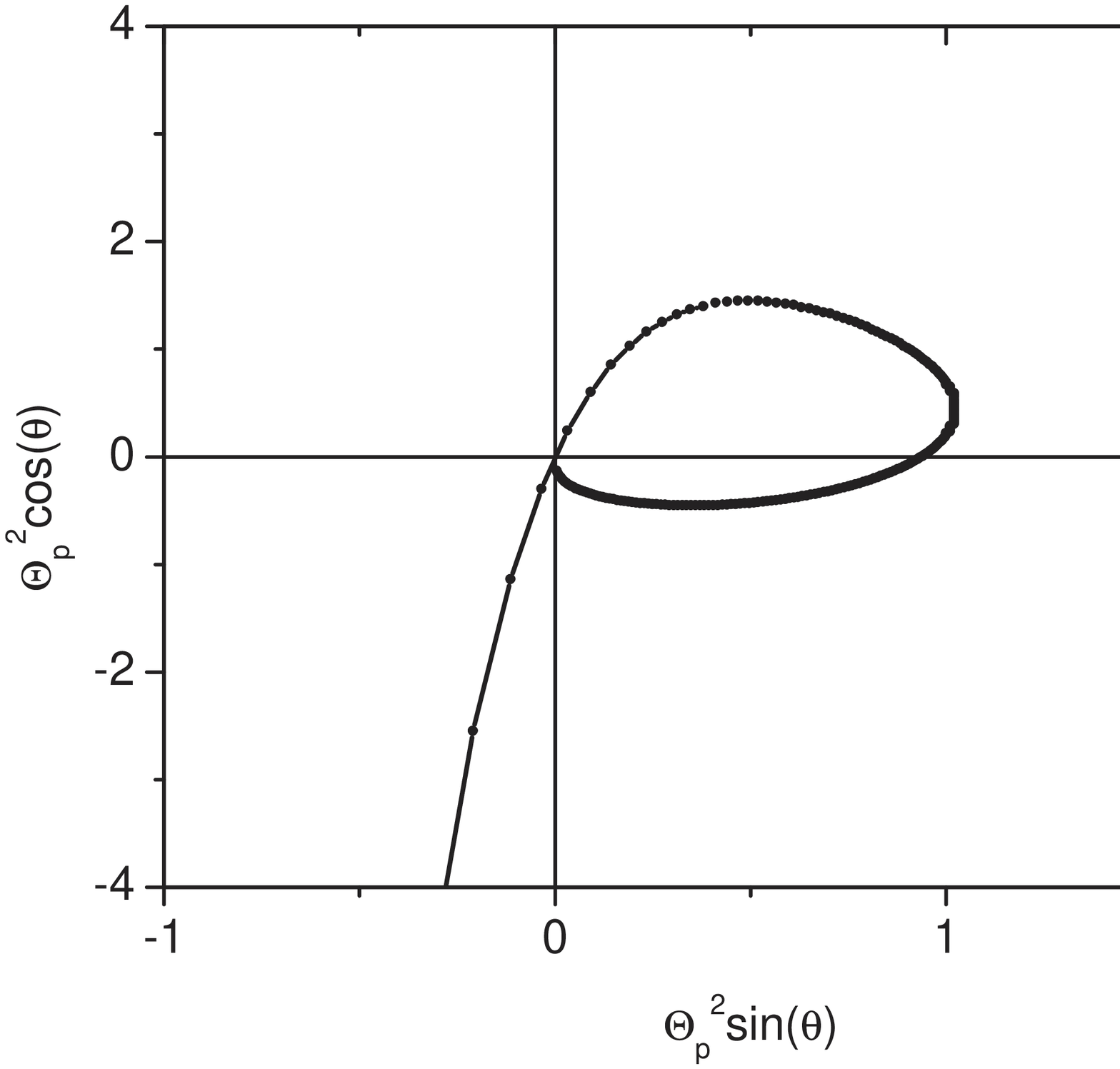}
\caption{The function $\Theta^{2}_{p}(x)$ with $x=\cos\theta$
 is shown in a polar plotted indicating a plasma pressure jet
 in a very narrow cone about $x=+1$.}
\label{fig.4}
\end{figure}

\clearpage
\begin{figure}
\plotone{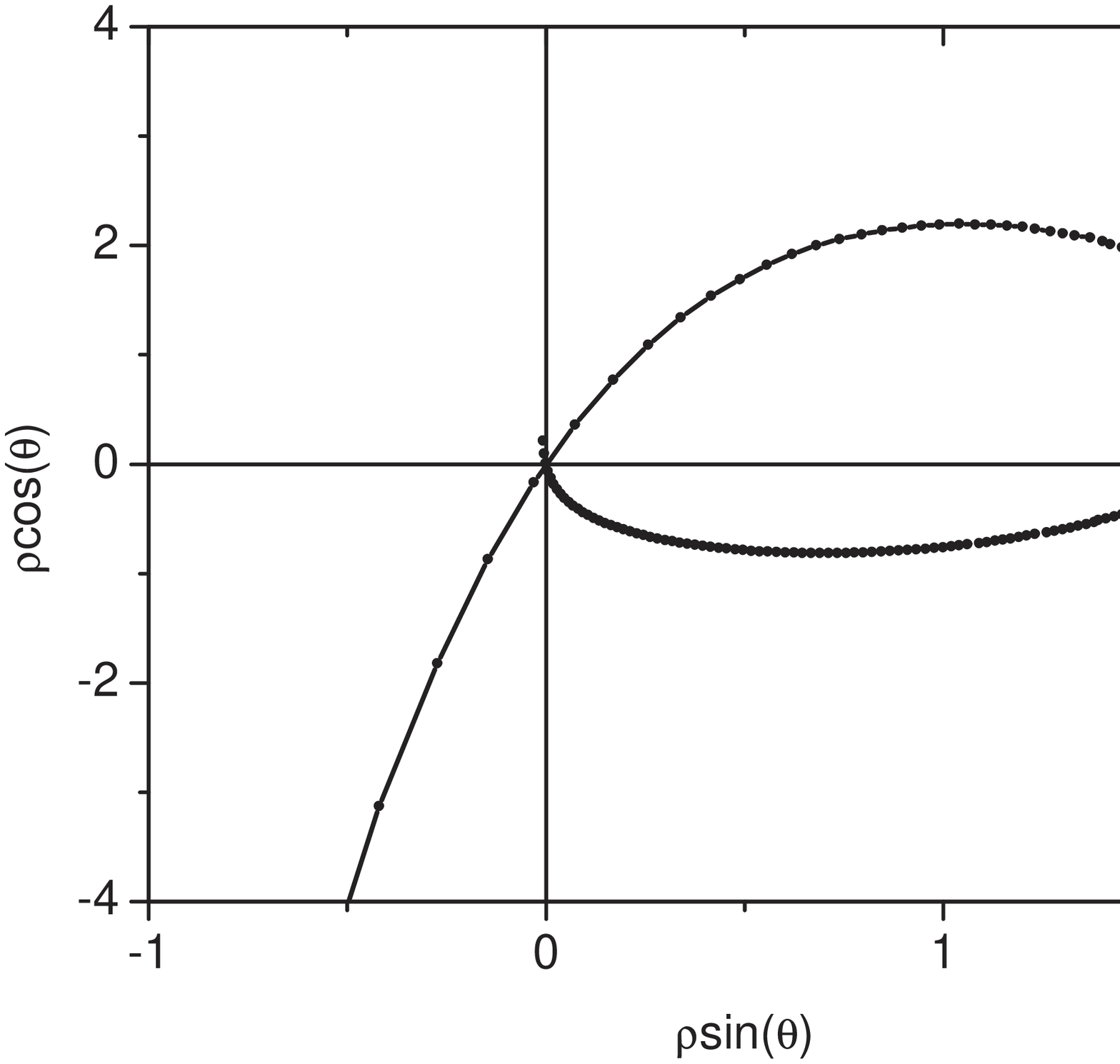}
\caption{The function $\tilde\rho(x)$ with $x=\cos\theta$
 is shown in a polar plotted indicating a mass density jet
 in a very narrow cone about $x=+1$.}
\label{fig.5}
\end{figure}

\clearpage
\begin{figure}
\plotone{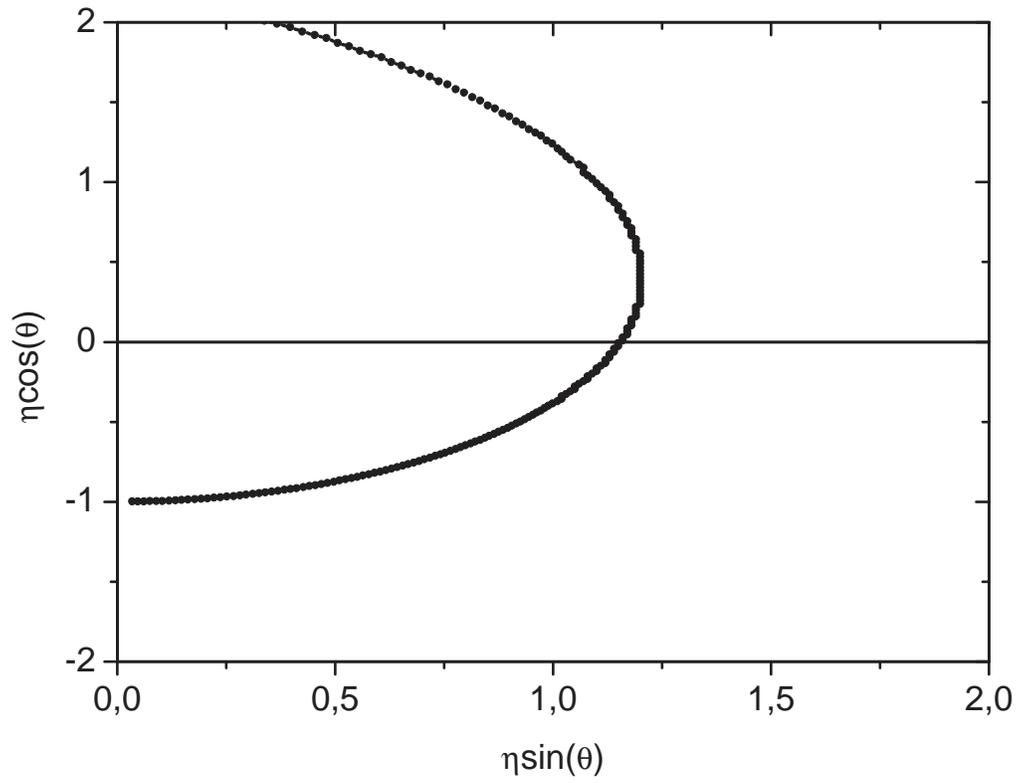}
\caption{The magnetic field line $\eta-\theta$ dependence
 is shown in a polar plotted.}
\label{fig.6}
\end{figure}

\clearpage
\begin{figure}
\plotone{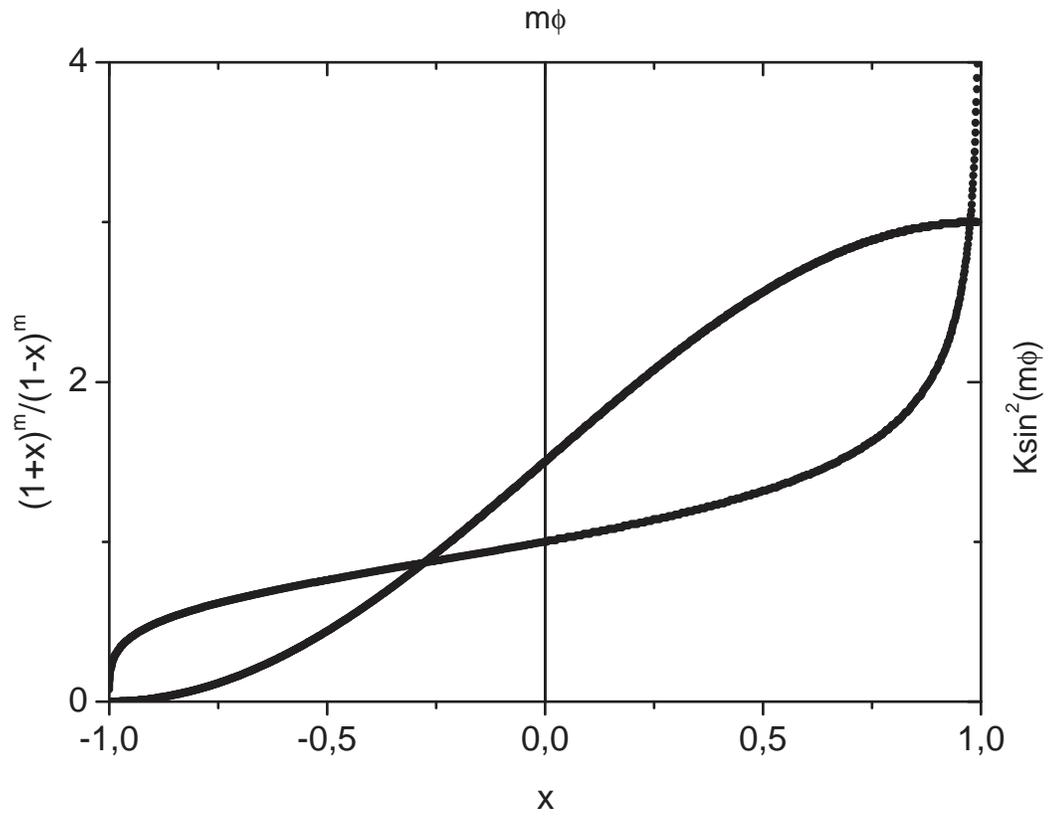}
\caption{The magnetic field line $\phi-\theta$ dependence
 is shown in a parametric plotted.}
\label{fig.7}
\end{figure}

\clearpage
\begin{figure}
\plotone{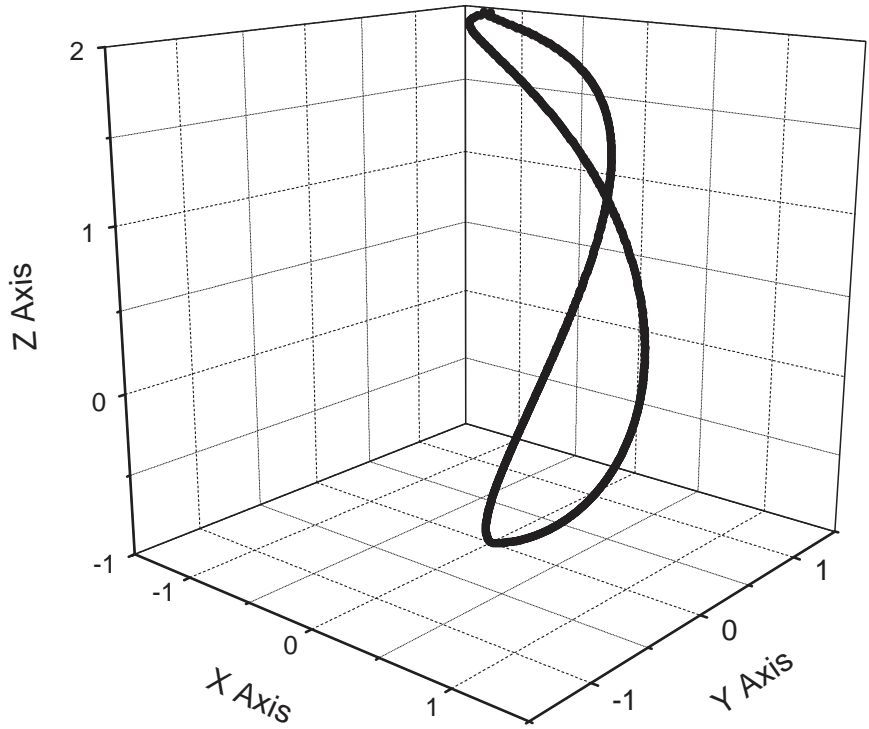}
\caption{The three-dimensional magnetic field lines,
 wounded on the surface of revolution of Fig.6,
 are viewed parallel to the x-y plane at about 45 degrees.}
\label{fig.8}
\end{figure}

\clearpage
\begin{figure}
\plotone{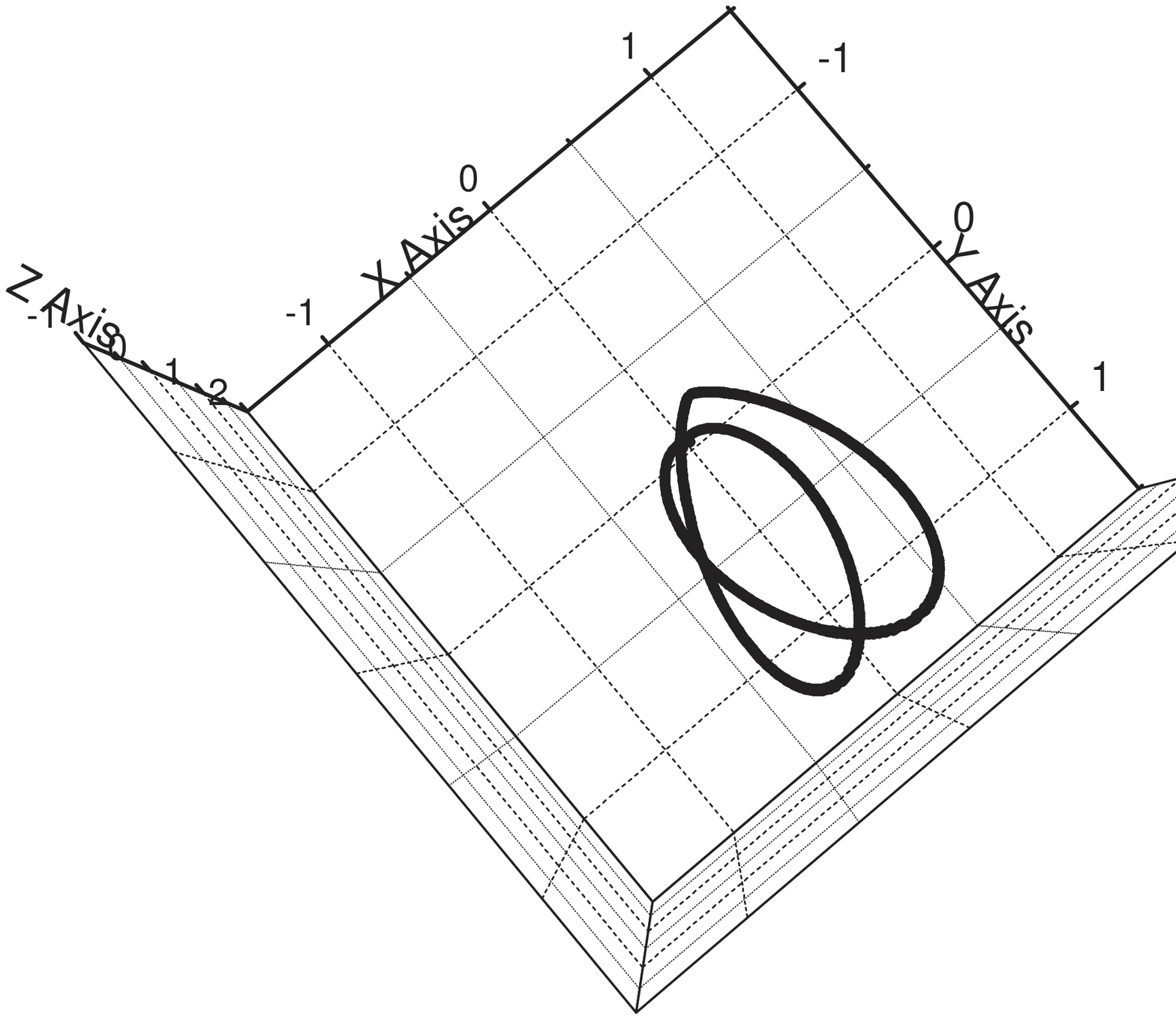}
\caption{The three-dimensional magnetic field lines,
 wounded on the surface of revolution of Fig.6,
 are viewed down the z axis.}
\label{fig.9}
\end{figure}

\clearpage
\begin{figure}
\plotone{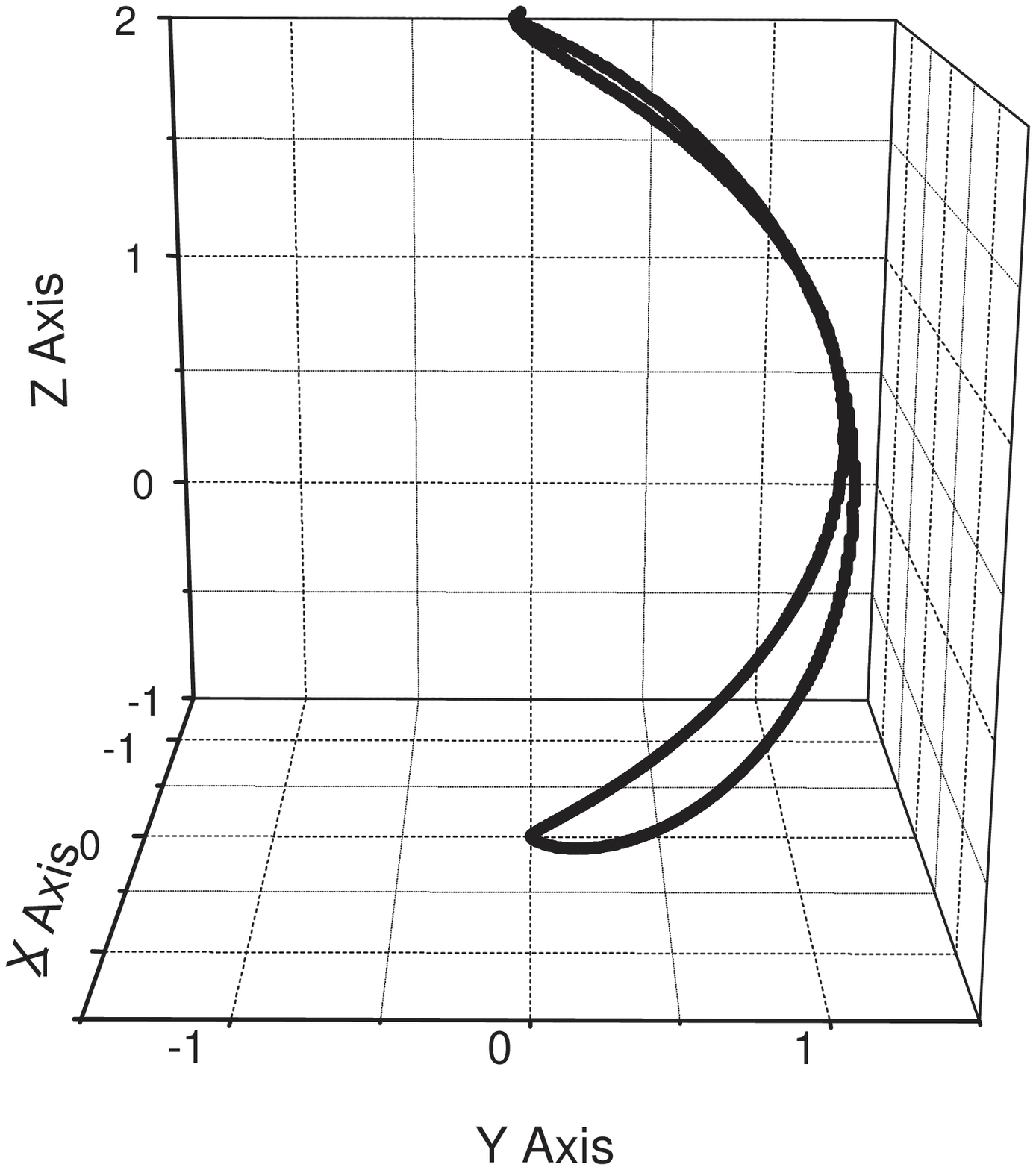}
\caption{The three-dimensional magnetic field lines,
 wounded on the surface of revolution of Fig.6,
 are viewed along the x axis.}
\label{fig.10}
\end{figure}

\clearpage
\begin{figure}
\plotone{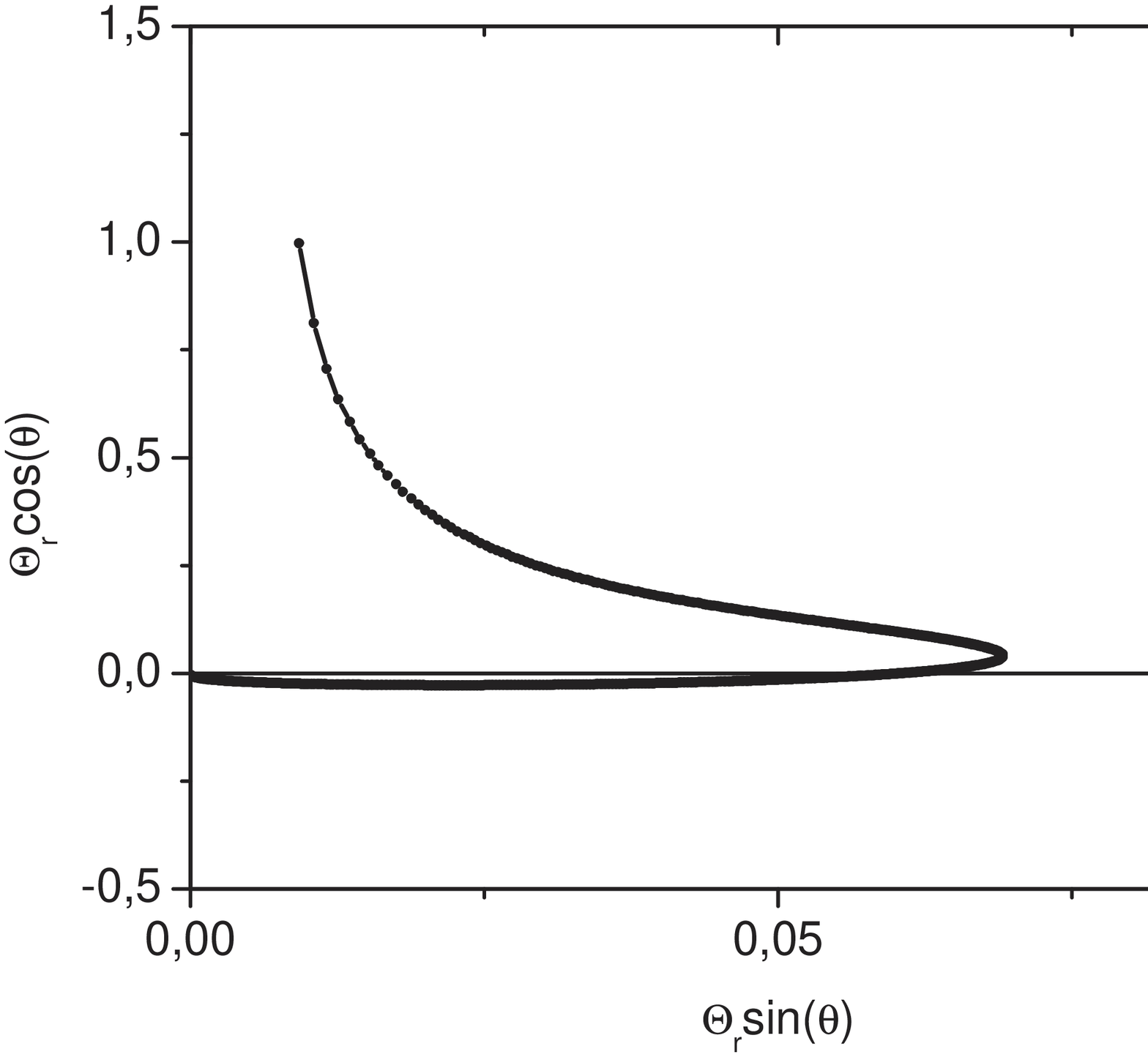}
\caption{The function $\Theta_{r}(x)$, with $n=3$,
 is shown in a polar plotted indicating an integrable singularity
 at $x=+1$ for a jet structure.}
\label{fig.11}
\end{figure}

\clearpage
\begin{figure}
\plotone{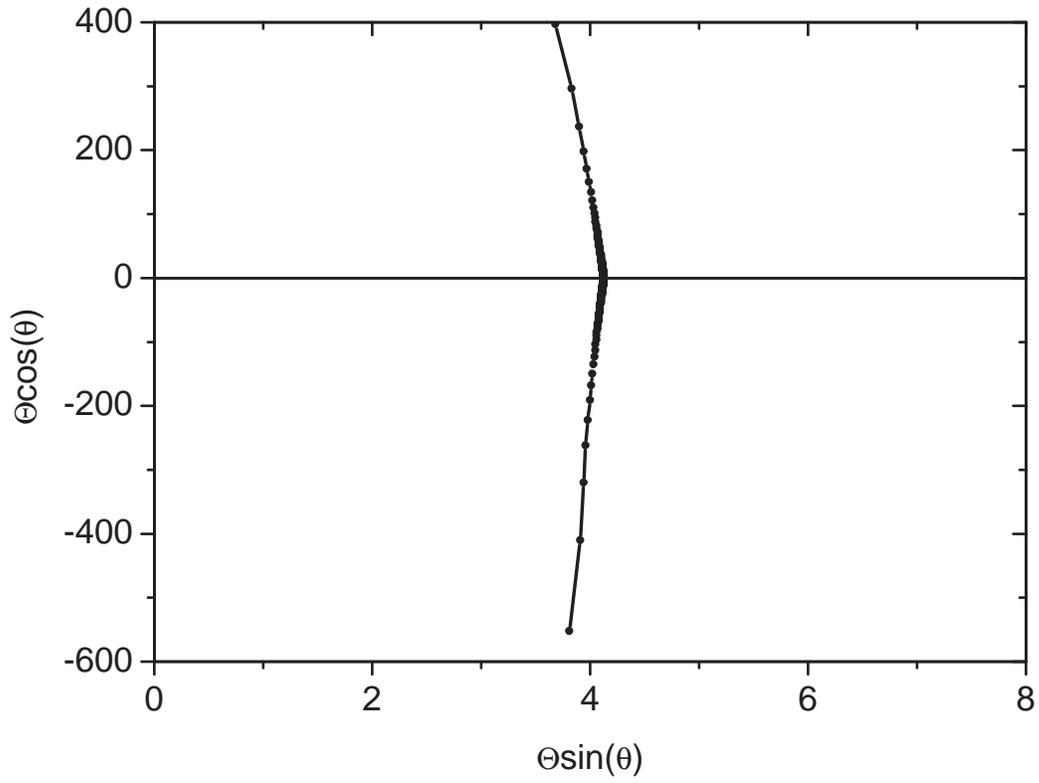}
\caption{The function $\Theta(x)$, with $n=3$,
 is shown in a polar plotted indicating a more symmetric structure.}
\label{fig.12}
\end{figure}

\clearpage
\begin{figure}
\plotone{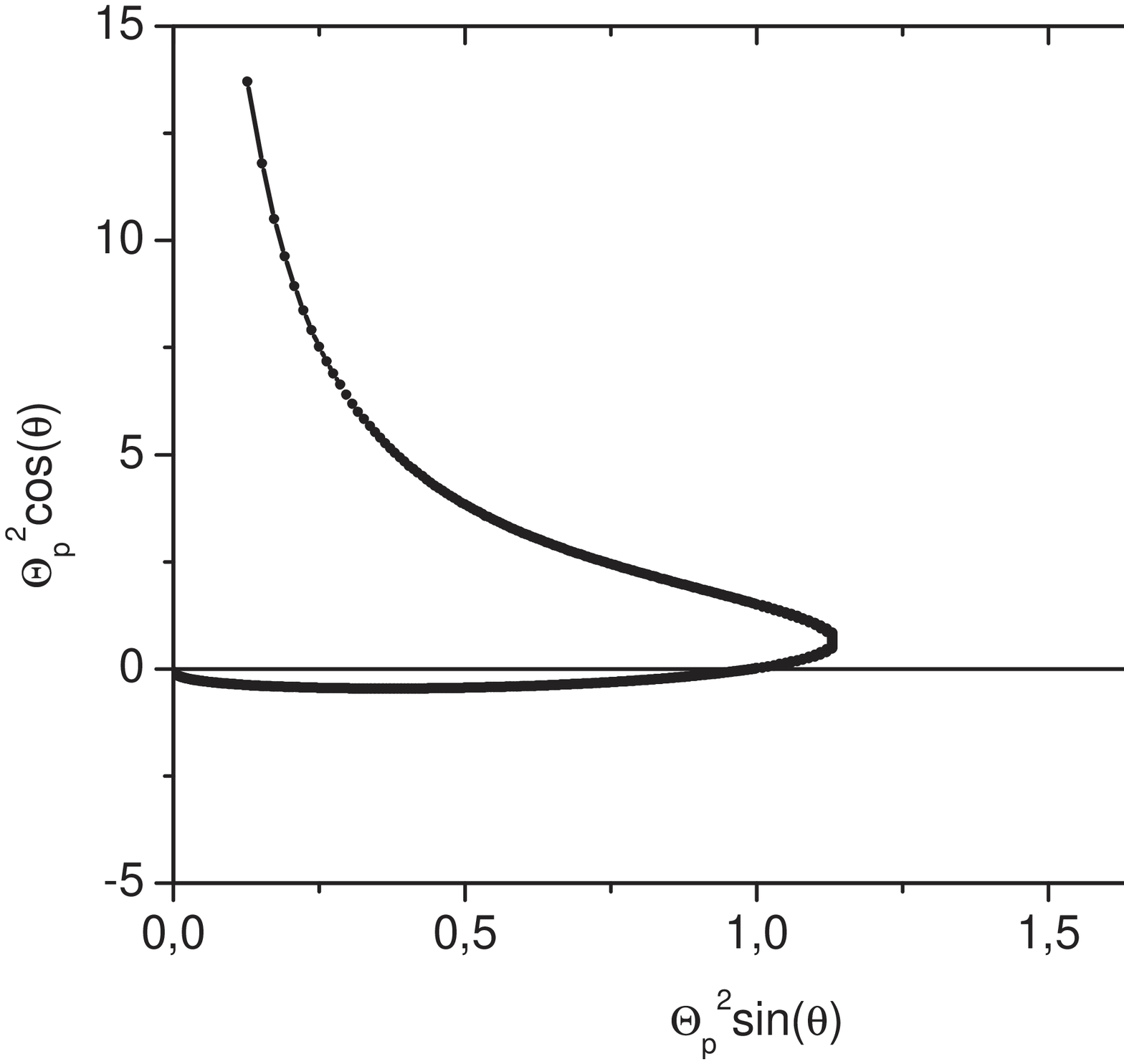}
\caption{The function $\Theta^{2}_{p}(x)$, with $n=3$,
 is shown in a polar plotted indicating a plasma pressure jet
 in a very narrow cone about $x=+1$.}
\label{fig.13}
\end{figure}

\clearpage
\begin{figure}
\plotone{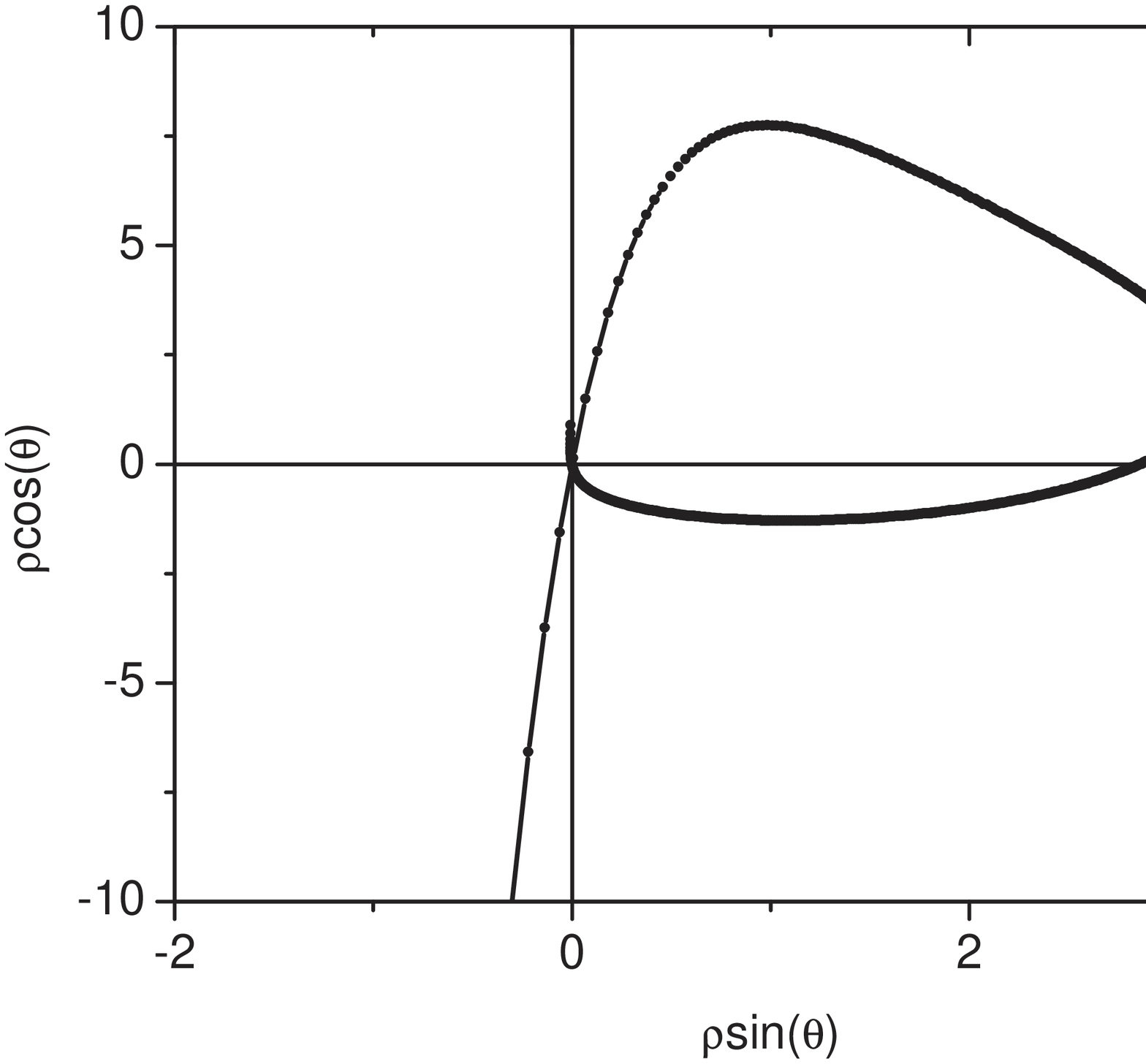}
\caption{The function $\tilde\rho(x)$ with $n=3$
 is shown in a polar plotted indicating a mass density jet
 in a very narrow cone about $x=+1$.}
\label{fig.14}
\end{figure}

\clearpage
\begin{figure}
\plotone{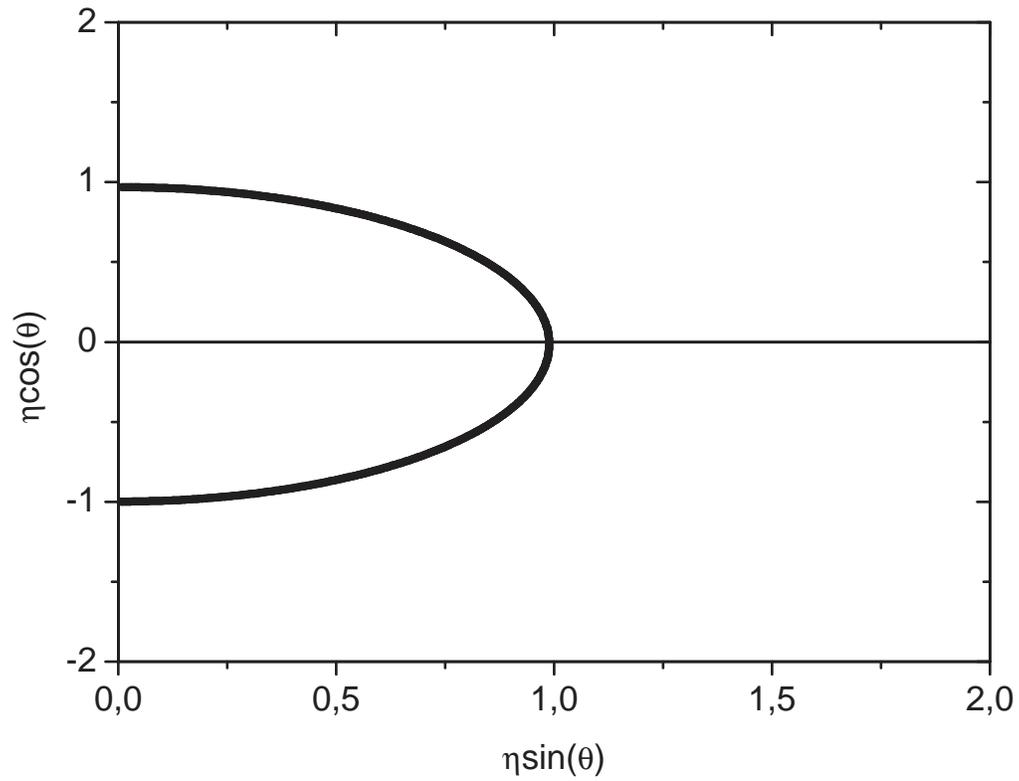}
\caption{The magnetic field line $\eta-\theta$ dependence,
 with $n=3$,
 is shown in a polar plotted
 showing a symmetric structure.}
\label{fig.15}
\end{figure}

\clearpage
\begin{figure}
\plotone{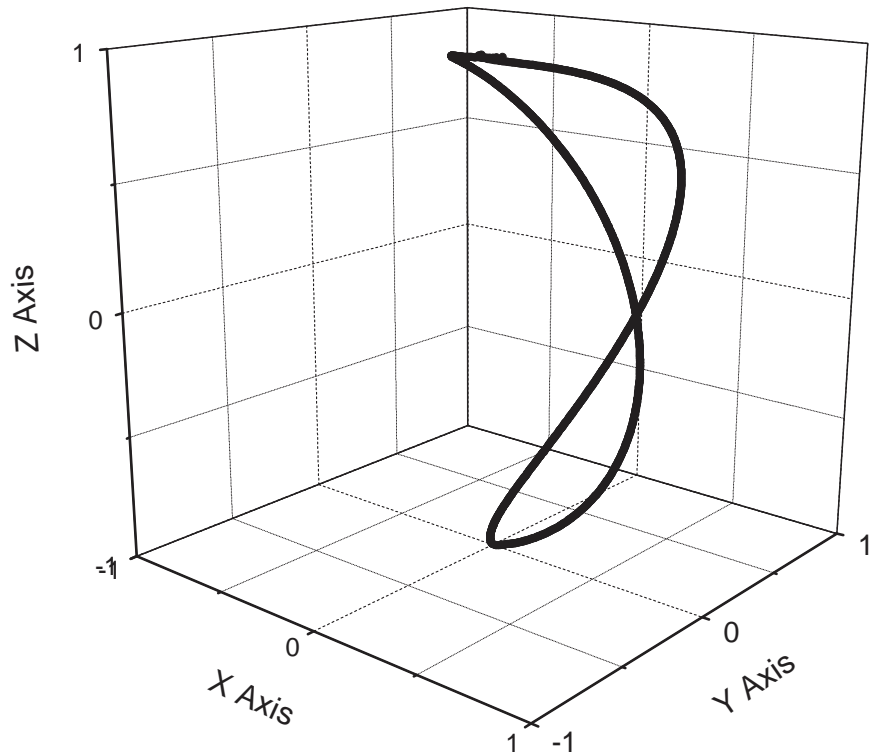}
\caption{The three-dimensional magnetic field lines,
 with $n=3$,
 are viewed parallel to the x-y plane at about 45 degrees.}
\label{fig.16}
\end{figure}

\clearpage
\begin{figure}
\plotone{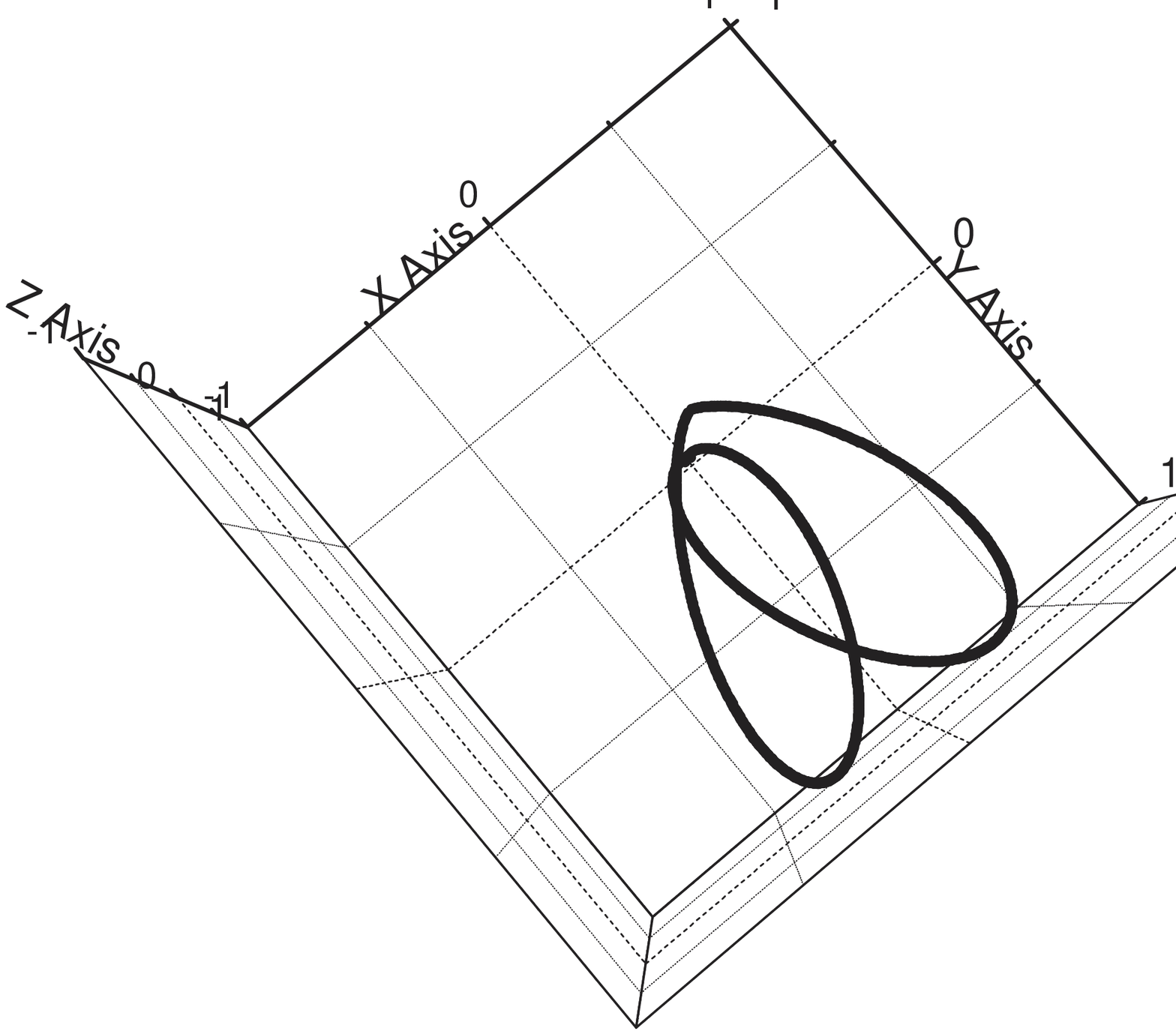}
\caption{The three-dimensional magnetic field lines,
 with $n=3$,
 are viewed down the z axis.}
\label{fig.17}
\end{figure}

\clearpage
\begin{figure}
\plotone{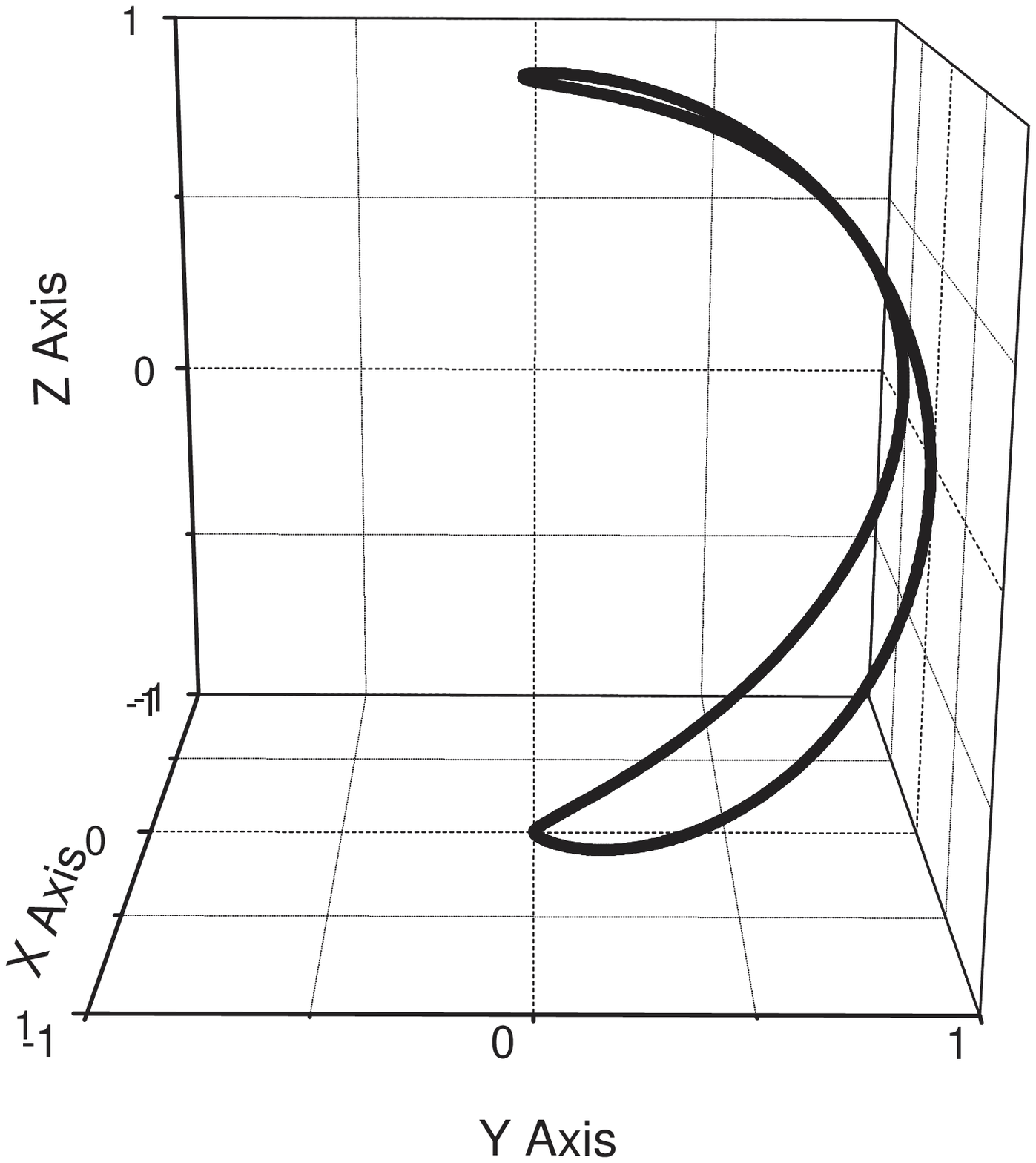}
\caption{The three-dimensional magnetic field lines,
 with $n=3$,
 are viewed along the x axis.}
\label{fig.18}
\end{figure}

\end{document}